\documentclass[
    ,final            
  ]
  {aipproc}

\layoutstyle{6x9}

\def\deg{$^\circ$}

\begin{document}

\title{X-ray pulsars through the eyes of INTEGRAL}


\author{Lutovinov Alexander, Tsygankov Sergey}{
  address={Space Research Institute, Profsoyuznaya str. 84/32, Moscow, Russia, 117997}
}



\begin{abstract}

Recent results of the spectral and timing analysis of X-ray pulsars in hard
X-rays with the INTEGRAL observatory are reviewed. The evolution of the
cyclotron line energy with the source luminosity was studied in detail for
the first time for several sources. It was shown that for V0332+53 this
dependence is linear, but for 4U0115+63 and A0535+262 it is more
complicated. There are some evidences of the ''reverse'' evolution for
GX301-2 and Her X-1, and no evolution was found for Vela X-1, Cen X-3,
etc. A strong dependence of the pulse fraction on the energy and source
luminosity was revealed and studied in detail. A prominent feature in the
pulse fraction dependence on the energy was discovered near the cyclotron
frequency for several bright sources. The obtained results are compared with
results of observations in standard X-rays and briefly discussed in terms of
current models; some preliminary explanations are proposed.

\end{abstract}

\maketitle


\section{Introduction}

According to the simple theory of accretion onto a rapidly rotating neutron
star with a strong magnetic field, the matter from the normal star is
stopped near the Alfven surface by the magnetic field pressure, trapped into
it and moved along the field lines to small ringing regions on the neutron
star surface, emitting X-rays. Due to the dipole structure of the neutron
star magnetic field two hot spots will be formed on the surface. In such a
case, for a rotating neutron star, the observer will detect on the source
light curve pulses of different forms depending on the physical and
geometrical conditions as near the neutron star surface, where these pulses
are formed, as on the line of sight. The corresponding pulse fraction
depends on the configuration of the emitted regions, relative position of
the dipole to the observer and the energy.

As it was shown in several early papers \cite{wangwel81,white83}, observed
pulse profiles are essentially differed for different sources; moreover,
they demonstrate the wide variety of the forms depending on the energy and
can shift, in some cases, up to 180\deg on the pulse phase. There is also a
strong pulse profile dependence on the source luminosity, which can be
connected with changes of the beam function from the pencil-beam to the
fan-beam ones with increasing luminosity from $<10^{37}$ to
$several\times10^{38}$ ergs s$^{-1}$ \cite{bs1976}. The presence of the
strong ($10^{11}-10^{13}$ G) magnetic field near the emitting regions on the
surface of the neutron star can be an origin of additional peculiarities in
the observed properties of X-ray pulsars, e.g. the presence of the cyclotron
resonance absorption features in the X-ray pulsar spectra, whose energy is
directly connected with the value of the magnetic field. Moreover,
properties of the accreting plasma are significantly changed at the
cyclotron frequency, that can leads to changes in the emission beam function
\cite{mesnag81}. The corresponding changes of the pulse profile near the
cyclotron energy were detected from several sources (see, e.g.,
\cite{tsy06}).

Obviously, such strong changes in the beam function, conditions and geometry
of the emission formation regions should lead to strong dependence of the
pulse fraction on the energy and source luminosity. This fact was recognized
as early as 80-90's, but only with launches of the orbital observatories
RXTE and INTEGRAL, which have high timing and energy resolution (especially
at high energies, where the observed emission isn't subject of the influence
of photoabsorption and depends only on the system geometry and physical
conditions in the formation regions) has it been possible to carry out
systematic investigations in this field. In particular, Tsygankov et
al. \cite{tsy07} showed that the pulse fraction of the X-ray pulsar
4U0115+634 is decreased with the source luminosity increase and increased
with the energy, having maximums near harmonics of the cyclotron absorption
line. The pulse fraction increase with the energy was also found for several
other X-ray pulsars (see, e.g.  \cite{fer07,barn08}). Using INTEGRAL data,
Tsygankov \& Lutovinov \cite{tsylut08} studied in detail the pulse fraction
dependence on the energy and luminosity for ten bright X-ray pulsars in hard
X-rays.

Here we summarize and briefly review current results of observations of
X-ray pulsars with the INTEGRAL observatory, drawing the main attention to
the cyclotron line energy, pulse profiles and pulse fraction dependences on
the luminosity and energy band.

\section{Spectra}

\begin{figure}
  \includegraphics[width=\textwidth,bb=35 145 560 720,clip]{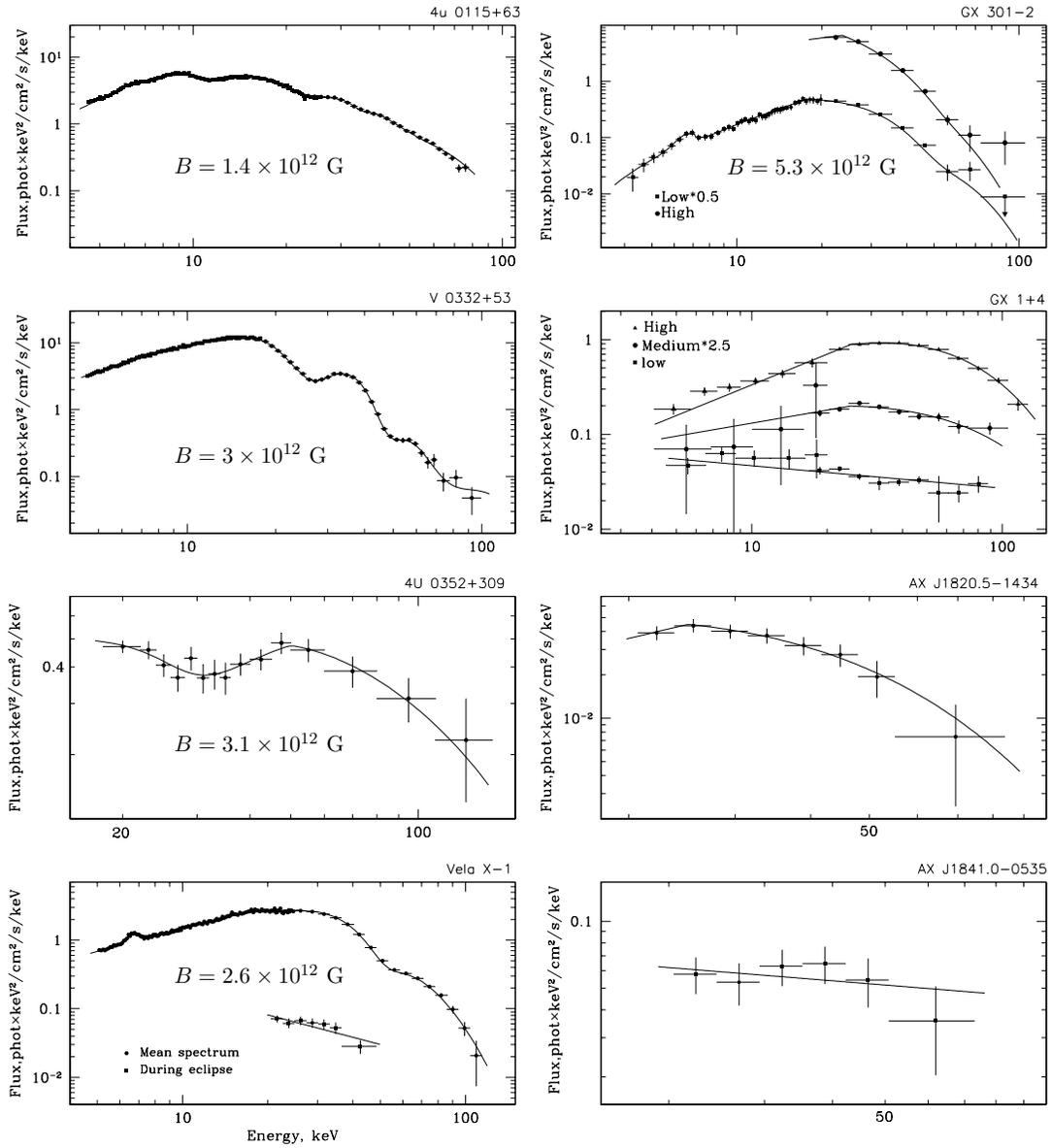}
  \caption{Spectra of eight X-ray pulsars obtained with the INTEGRAL
  observatory. Values of the magnetic field on the surface of the
  neutron star are indicating for sources with detecting cyclotron
  lines in their spectra.}\label{spectra}
\end{figure}

The INTEGRAL observatory \cite{win03} has observed more than 70 X-ray
pulsars (including several new ones) during $\sim6$ years of operation in
orbit; 38 of them were detected at a high significance level, that gave us a
possibility to reconstruct their spectra in hard X-rays. For several sources
it was possible to perform a search for cyclotron absorption features and
carry out detailed analysis of the spectral variability depending on its
luminosity (see, e.g., \cite{fil05}). The list of bright X-ray pulsars
($>100$ mCrab in the $20-100$ keV energy band) with corresponding energies
of the cyclotron line and its harmonics is presented below:

\vspace{1mm}

\begin{tabular}{ll}

4U 0115+63   &  $\sim$11, 22, 34, 44 keV \\
V 0332+53    &  $\sim$28, 55 keV \\
A 0535+262   &  $\sim$45, 100 keV \\
Her X-1      &  $\sim$38 keV \\
Vela X-1     &  $\sim$26, 56 keV \\
GX 301-2     &  $\sim$50 keV \\
Cen X-3      &  $\sim$31 keV \\
4U 0352+30$^{*}$   &  $\sim$30 keV \\
GX 1+4       &   --$^{a}$    \\
OAO 1657-415 &   --    \\
EXO 2030+375 &   --    \\

\end{tabular}
\\
\begin{footnotesize}
\begin{tabular}{cl}
$^{*}$ -- & maximum source flux was $\sim40$ mCrab, but the cyclotron line was detected in its spectrum (\cite{lut04}) \\
$^{a}$ -- & no cyclotron line detected with INTEGRAL \\
\end{tabular}
\end{footnotesize}
In Fig.\ref{spectra} the spectra of eight X-ray pulsars of the different
nature are presented for illustration: two of these objects (AX J1820.5-1434
and AX J1841.0-0535) were registered in hard X-rays for the first time; in
spectra of several other sources the cyclotron absorption features were
clearly detected; strong variability of the spectral shape of the X-ray
pulsars GX1+4 and GX301-2 with the luminosity was observed. Note the first
detection of the hard X-ray emission from Vela X-1 during the eclipse and
its spectrum reconstruction.

\begin{figure}
\hbox{
  \includegraphics[width=0.5\textwidth,bb=17 275 515 690,clip]{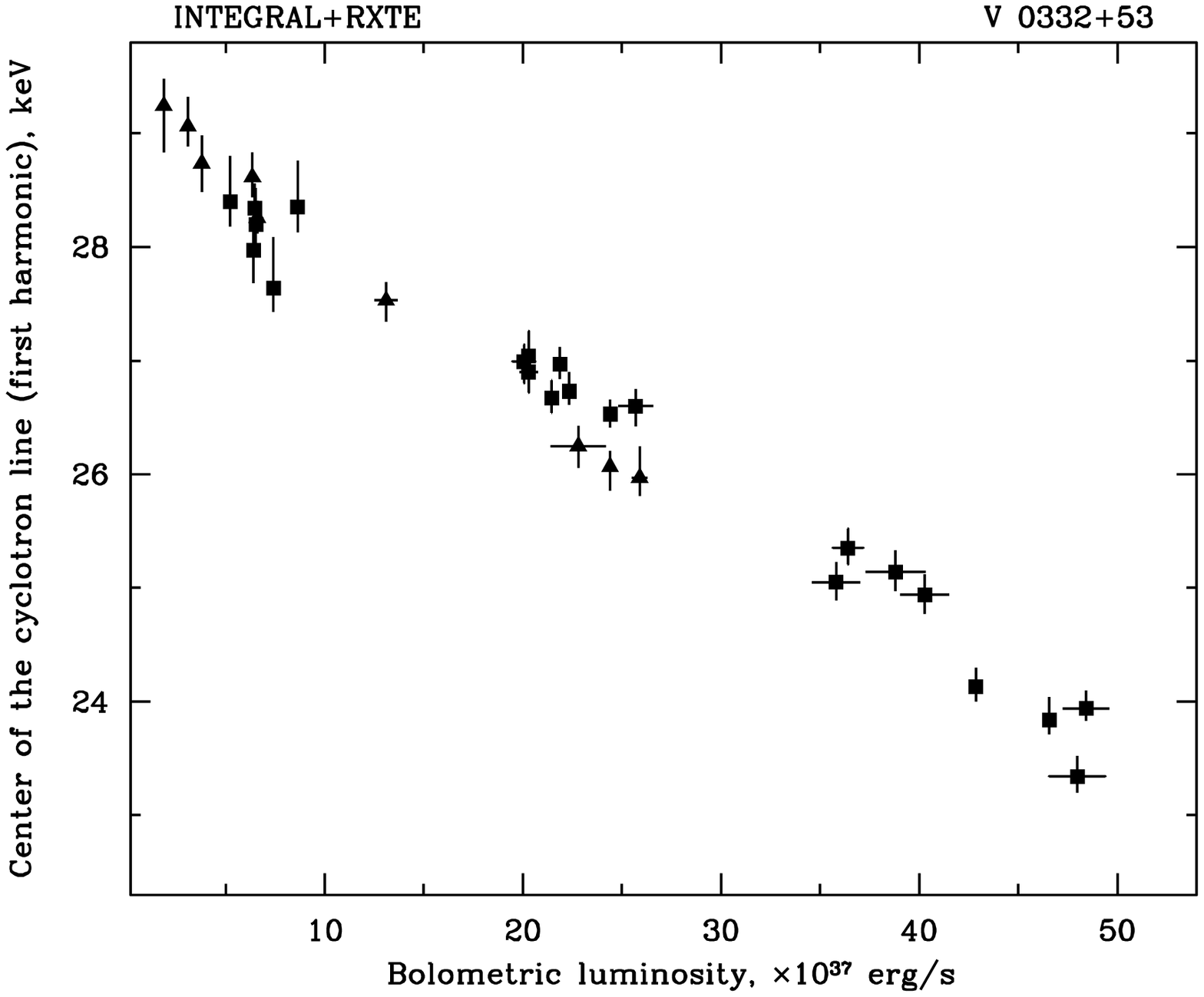}
  \includegraphics[width=0.5\textwidth,bb=17 275 515 690,clip]{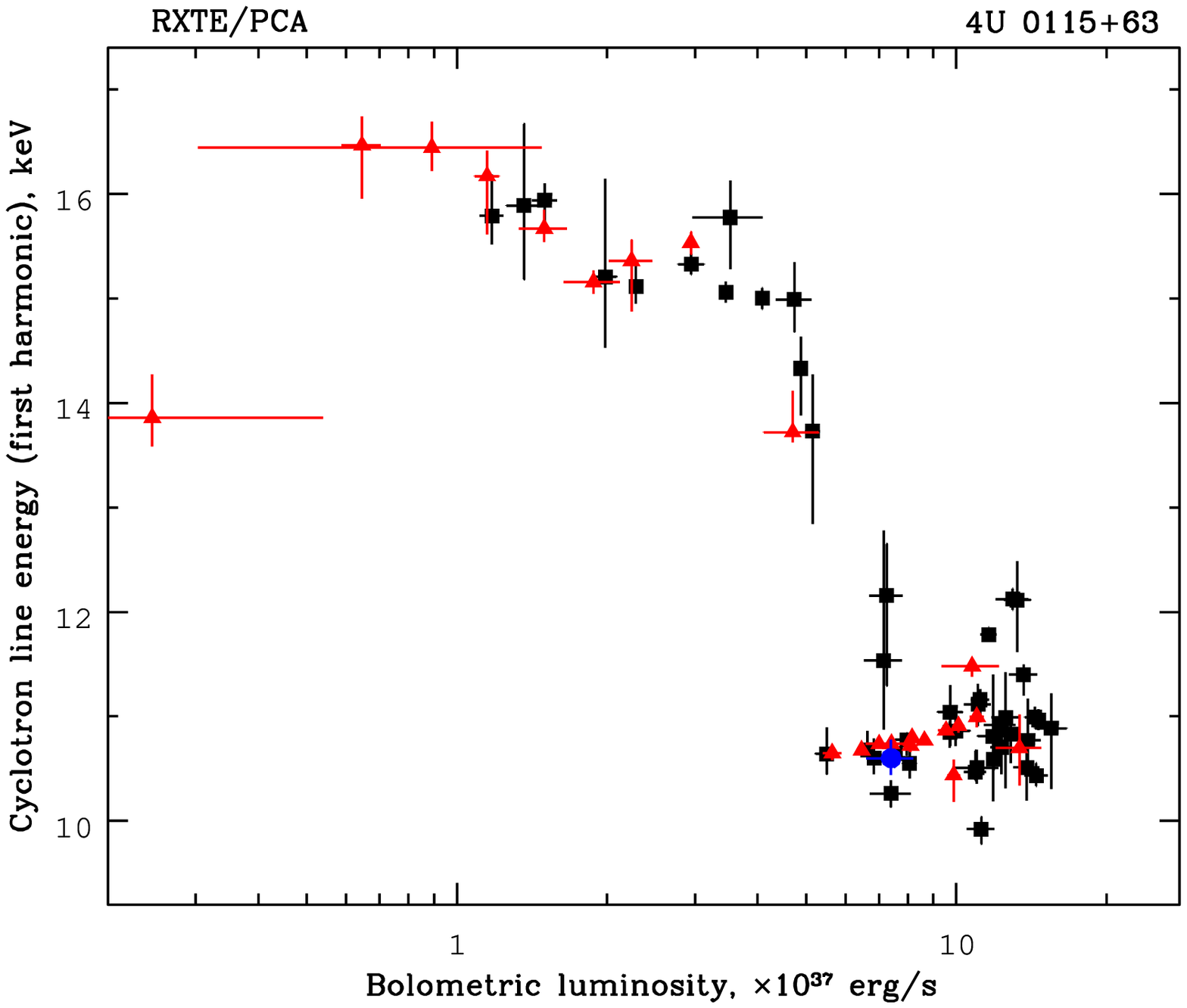}
}
  \caption{(a) The cyclotron line energy dependence on the source luminosity
  (3-100 keV) for V0332+52. Triangles are INTEGRAL results, squares are RXTE
  ones. (b) The same, but for 4U0115+63. Squares are RXTE measuremets
  performed during outburst of 1999, triangles are RXTE measurements during
  outburst of 2004, a circle is the INTEGRAL measurement. }\label{cycline}
\end{figure}

For several X-ray pulsars it was known before that the energy of the
cyclotron line can somehow depend on the source luminosity (see, e.g.,
\cite{mih98}). Using INTEGRAL and RXTE data Tysgankov et al. \cite{tsy06}
showed for the first time that the energy of the main harmonic of the
cyclotron line in the V0332+53 spectrum grows approximately linearly with
the decrease of the source luminosity.  Such a behaviour was predicted by
theoreticians about 30 year ago, and now it was firstly observed (see
Fig.\ref{cycline}a).  In approaching the dipole magnetic field of the
neutron star, the maximum measured change of the cyclotron line energy
corresponds to $\sim750$ m of the relative change of the height $h$ above
the neutron star surface. This height may be considered as an averaged or
``effective'' height of the formation of the cyclotron feature
\cite{tsy06}. Moreover, for the first time it was shown that the behaviour
of the second harmonics energy is qualitatively similar to the main
one. Altogether in a high luminosity state 3 harmonics of the cyclotron line
were detected (see Fig.\ref{spectra}).

Another well known transient X-ray pulsar 4U 0115+63 was analysed with
INTEGRAL and RXTE observatories during its intense outbursts in 1999 and
2004 \cite{tsy07}. Due to the high source brightness and relatively low
magnetic field of the neutron star four harmonics of the cyclotron line were
detected near $\sim$ 11, 22, 33 and 44 keV.  But, in contrast to V0332+53,
the cyclotron energy dependence on the pulsars' luminosity could not be
approximated by a simple linear law in a wide luminosity range. In the high
luminosity state ($5\times10^{37}$ -- $2\times10^{38}$ erg s$^{-1}$) the
energy of the fundamental harmonic is practically constant ($\sim$ 11 keV);
when the pulsar luminosity falls below $\sim5\times10^{37}$ erg s$^{-1}$,
the energy of the fundamental harmonic is displaced sharply toward the high
energies, up to $\sim16$ keV (Fig.\ref{cycline}b). Under the assumption of a
dipole magnetic field configuration, this change in the cyclotron energy
corresponds to the decrease of the height of the emitting region by
$\sim2$~km. Note, that such a behaviour was observed during both outburst,
indicating that it can be a fundamental property of the pulsar.

Similar investigations were performed for most of bright X-ray pulsars with
cyclotron lines and was found that: the complex behaviour of the cyclotron
line energy was detected in the spectrum of A0535+262 during the powerful
outburst \cite{cab08}; a positive correlation between the cyclotron line
energy and source luminosity was found for Her X-1, contrary to what is
observed in the transient pulsar V0332+53 \cite{staub07}; some evidences of
similar behaviour was registered also for GX301-2 \cite{fil05}; no evolution
of the cyclotron line was found for other sources, like Vela X-1, Cen X-3,
etc.

An interesting feature is a nonequidistance of the cyclotron line harmonics,
which was observed in spectra of X-ray pulsars (see table above). The theory
predicts that the accretion column emits harder radiation from regions
closer to the neutron star surface, where the magnetic field strength is
higher. Thus, we assumed that deviations of energies of the fundamental and
higher harmonics in the spectrum of 4U0115+63 from a linear law can be used
to compare the effective sizes of the emitting regions at $\sim$11, 22, 33,
and 44 keV, respectively. Due to a higher magnetic field near the neutron
star surface the energies of the higher harmonics should lie above the
harmonic law. At the luminosity of $\sim7\times10^{37}$ erg s$^{-1}$, four
almost equidistant cyclotron line harmonics were clearly detected in the
spectrum of 4U0115+63. This suggests that either the region where the
emission originates is compact or the emergent spectrum from different (in
height) segments of the accretion column is uniform.  At higher luminosity
($\sim11\times10^{37}$ erg s$^{-1}$), where the strong nonequidistance was
observed, the possible scatter of heights, at which the emission with
different energies is produced, was found to be comparable with the height
variability obtained from variations of fundamental harmonic energy with the
luminosity \cite{tsy07}.

\section{Timing analysis}

\begin{figure}
  \includegraphics[width=0.95\textwidth,bb=0 100 580 810,clip]{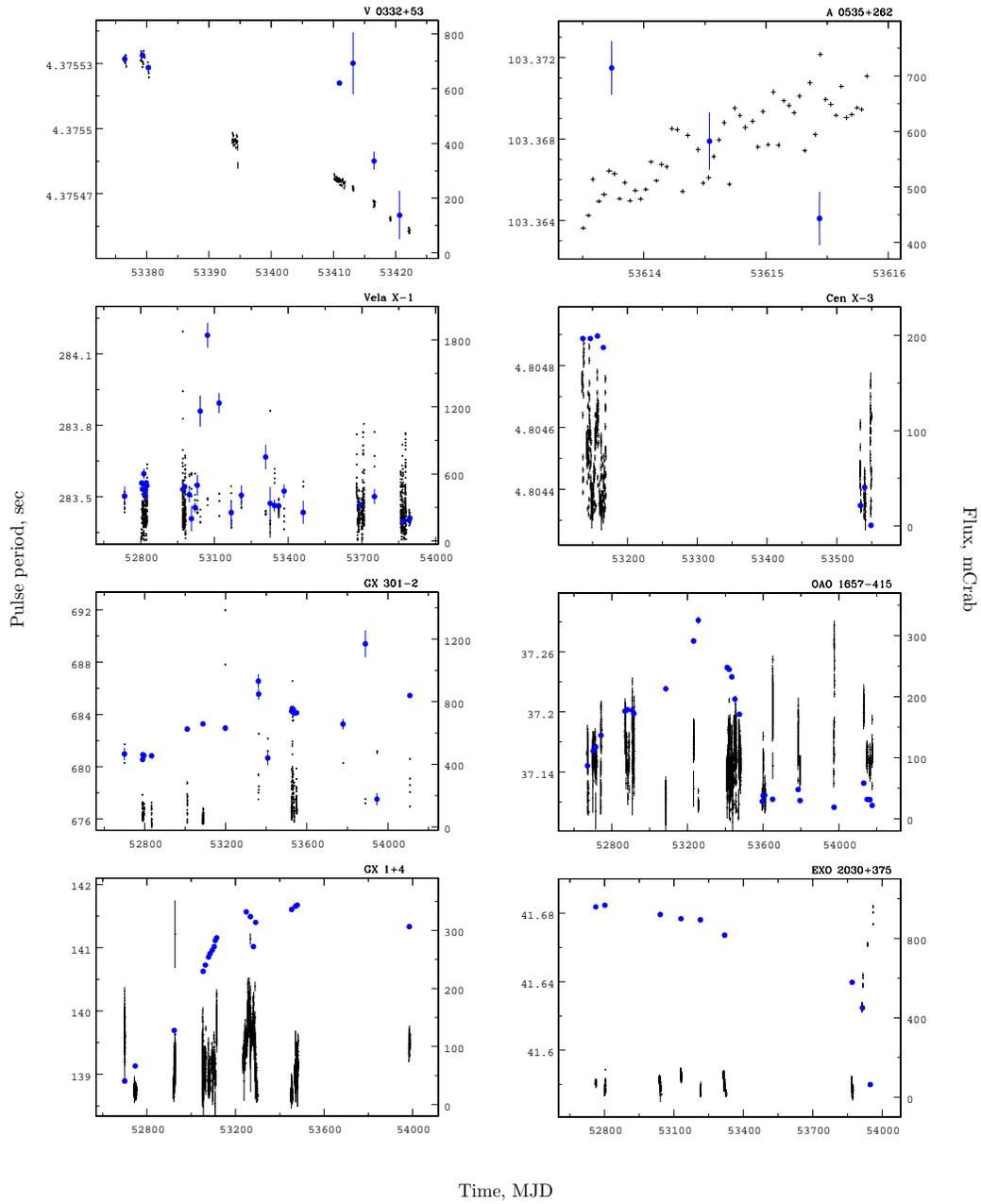}
  \caption{Light curves of X-ray pulsars in the 20-60 keV energy band (crosses) and 
histories of pulse periods (circles), as measured by INTEGRAL.}\label{periods}
\end{figure}

We performed an analysis of bright X-ray pulsars to search pecularities of
hard X-ray emission at different time scales and to trace changes of its
timing properties. As a result of this analysis a number of pulse profiles
in several energy bands, 2D-maps of the emission intensity in the
coordinates of energy and pulse phase and pulse fraction (calculating as
$PF=\frac{I_{max}-I_{min}}{I_{max}+I_{min}}$, where $I_{max}$ and $I_{min}$
are the maximum and minimum intensities of the pulse profile) dependence on
the energy were obtained for each of pulsars. The description of applied
methods and technique can be found in \cite{tsy06,tsylut08}. In
Fig.\ref{periods} the pulse period histories and corresponding light curves
in the 20-60 keV energy band, obtained during INTEGRAL observations, are
presented for several of the pulsars. It is clearly seen that for some of
pulsars there is a clear correlation between the source flux and pulse
period, especially during the outbursts, when the accretion rate onto the
neutron star is increased significantly. Such a correlation can be roughly
explained within the theory of magnetic moment transfer \cite{ghosh79},
which connects the neutron star magnetic moment, the pulse period and its
change, source flux and distance to the system. To take an illustration,
applying this theory to the X-ray pulsar A0535+262 we can obtain the
estimation of the distance to the system, $3.1\pm0.7$ kpc, that is in
agreement with results of optical observations (\cite{pacheco87}).

To check the accuracy and correctness of the performed analysis we compared
results of observations of the Crab pulsar by the INTEGRAL and Jodrell Bank
observatories in hard X-ray and radio bands, respectively, and found that
with high accuracy (a typical difference $\sim1\times10^{-9}$ s) they
agree well.

\subsection{Pulse Profiles}

\begin{figure}
  \includegraphics[height=1.02\textwidth,bb=65 65 570 780,clip,angle=90]{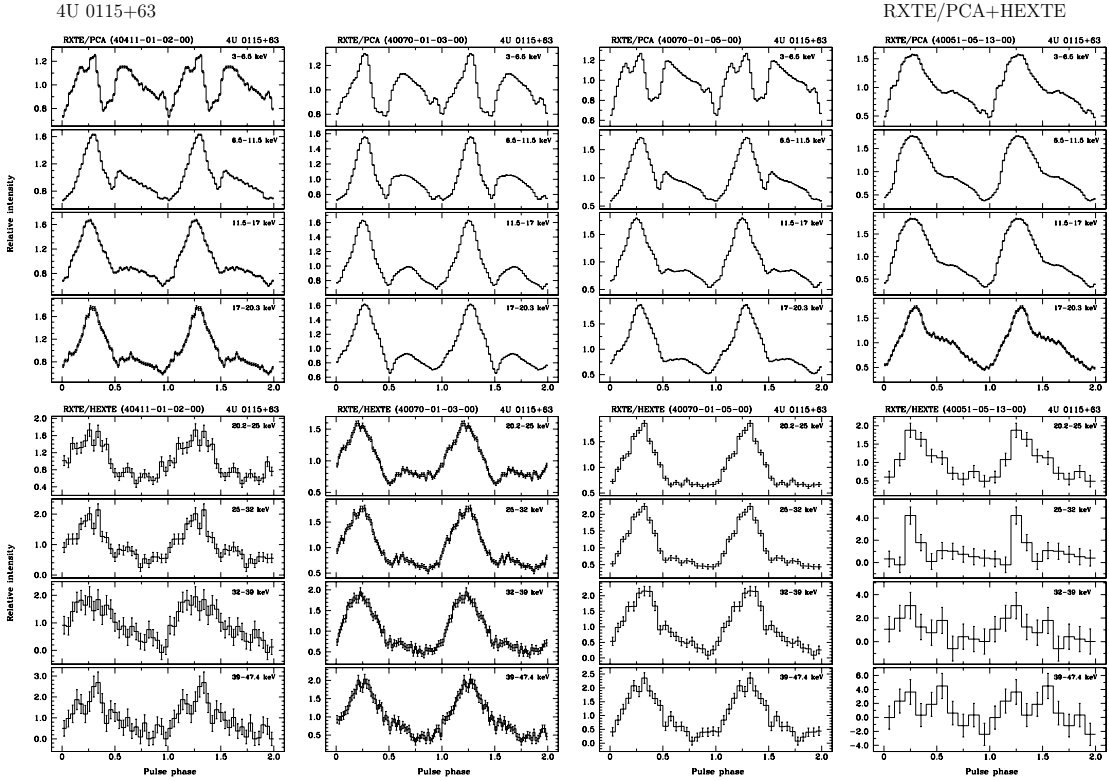}
  \caption{Evolution of the 4U0115+63 pulse profile with the energy and 
source luminosity; from the left to the right -- $7.3\times10^{37}$,
$14.6\times10^{37}$, $6.6\times10^{37}$, $1.5\times10^{37}$ ergs s$^{-1}$ 
(from \cite{tsy07}).}\label{pp0115}
\end{figure}

In spite of a few decades of observations there is no self consistent model
of pulse profile formation still proposed. Therefore it was very important
to find out some new features responsible for specific physical processes in
the accretional region near the neutron star surface which should be taken
into account for correct simulations of pulse profiles.  Particularly, for
several of the brightest X-ray pulsars it was possible for the first time to
study in hard X-rays the pulse profile changes depending on the energy and
source luminosity and compare them with results in standard X-rays. As an
example of such an analysis Fig.\ref{pp0115} shows the background-corrected
pulse profiles of 4U0115+63 at different source luminosities in different
energy bands. It is interesting to trace the evolution of the pulse profile
with energy and luminosity: at the soft energies the profile is
double-peaked with a tendency for the second peak to disappear with the
pulsar luminosity decreases; as the energy increases, the second peak also
disappears and the profile becomes virtually single-peaked above 20 keV.

\begin{figure}
  \includegraphics[width=6cm]{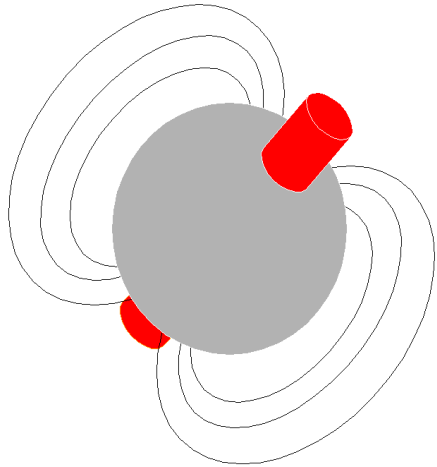}
  \caption{The geometrical "toy" model of the X-ray pulsar, taking into
  account the relative positions of accretion columns, neutron star and
  observer.}\label{toy}
\end{figure}

The decrease in the intensity of the second peak with decreasing luminosity
and increasing energy can be qualitatively explained by a simple, purely
geometrical model that is capable of describing the main observed trends in
the pulse profile in general terms (see Fig.\ref{toy}).  The rotation axis
of the neutron star is inclined with respect to its magnetic field axes in
such a way that the accretion column at one of the poles is seen over its
entire (or almost entire) height when the pole falls on the observer's line
of sight. Only the upper part emitting softer photons is seen in the second
column, while the emission region of hard photons is screened by the
neutron-star surface (hence the observed decrease in the intensity of the
second peak with increasing energy); as the accretion rate and, accordingly,
the source luminosity decrease, the column height decreases, the intensity
of the second peak falls, and we will cease to see it altogether at some
time.

Naturally, this is only a simple ''toy'' model and, to describe the observed
behavior of the pulse profiles more or less accurately, we must include to
the model the temperature distribution along the accretion column, the shape
of the beam function, its dependence on the object's luminosity and the
energy band, the light bending, etc.

\begin{figure}
\vbox{
\hbox{
\hspace{-6mm} \includegraphics[width=0.54\textwidth,bb=65 206 493 710,clip]{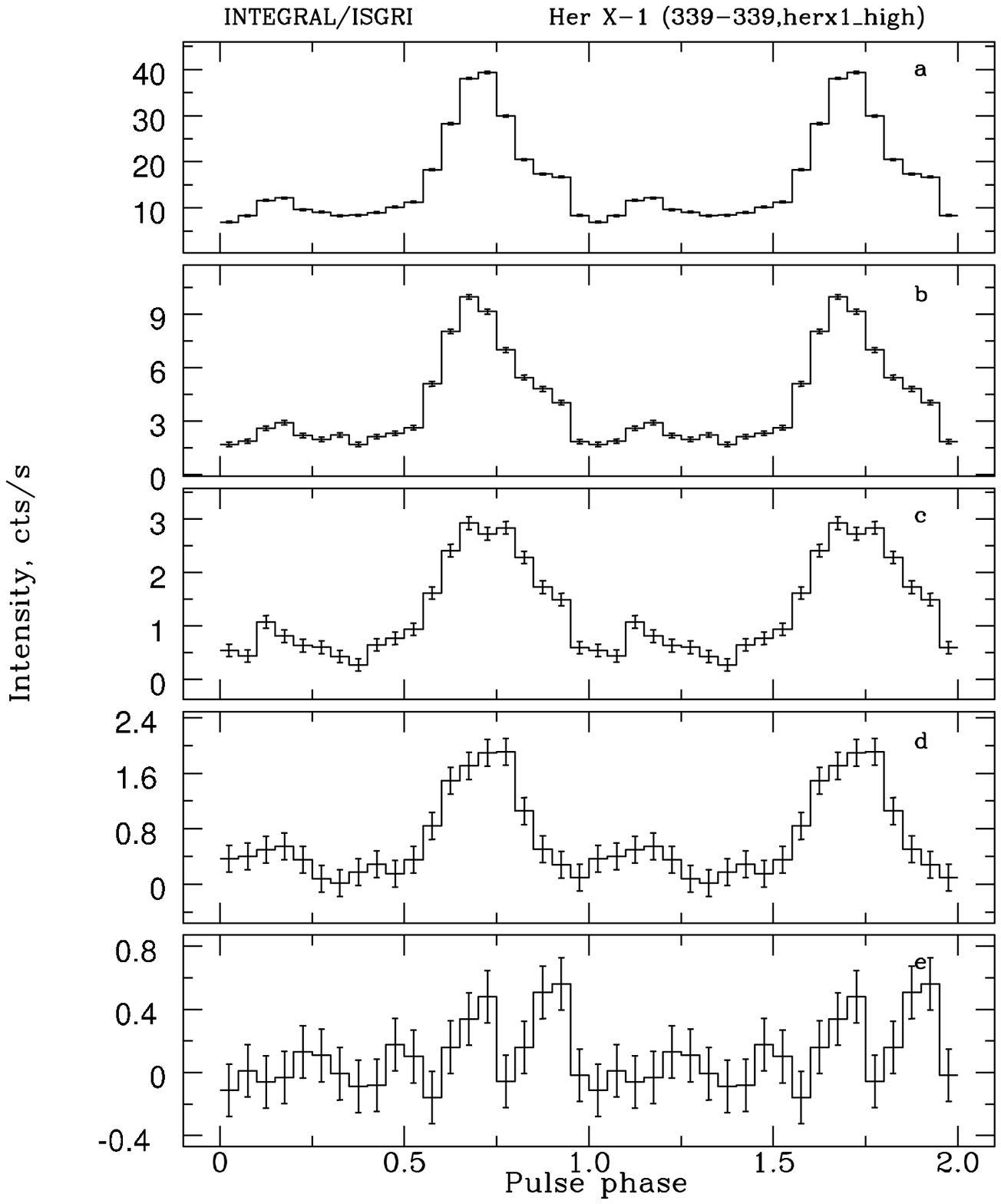}
\hspace{-6mm}  \includegraphics[width=0.52\textwidth,bb=80 206 493 710,clip]{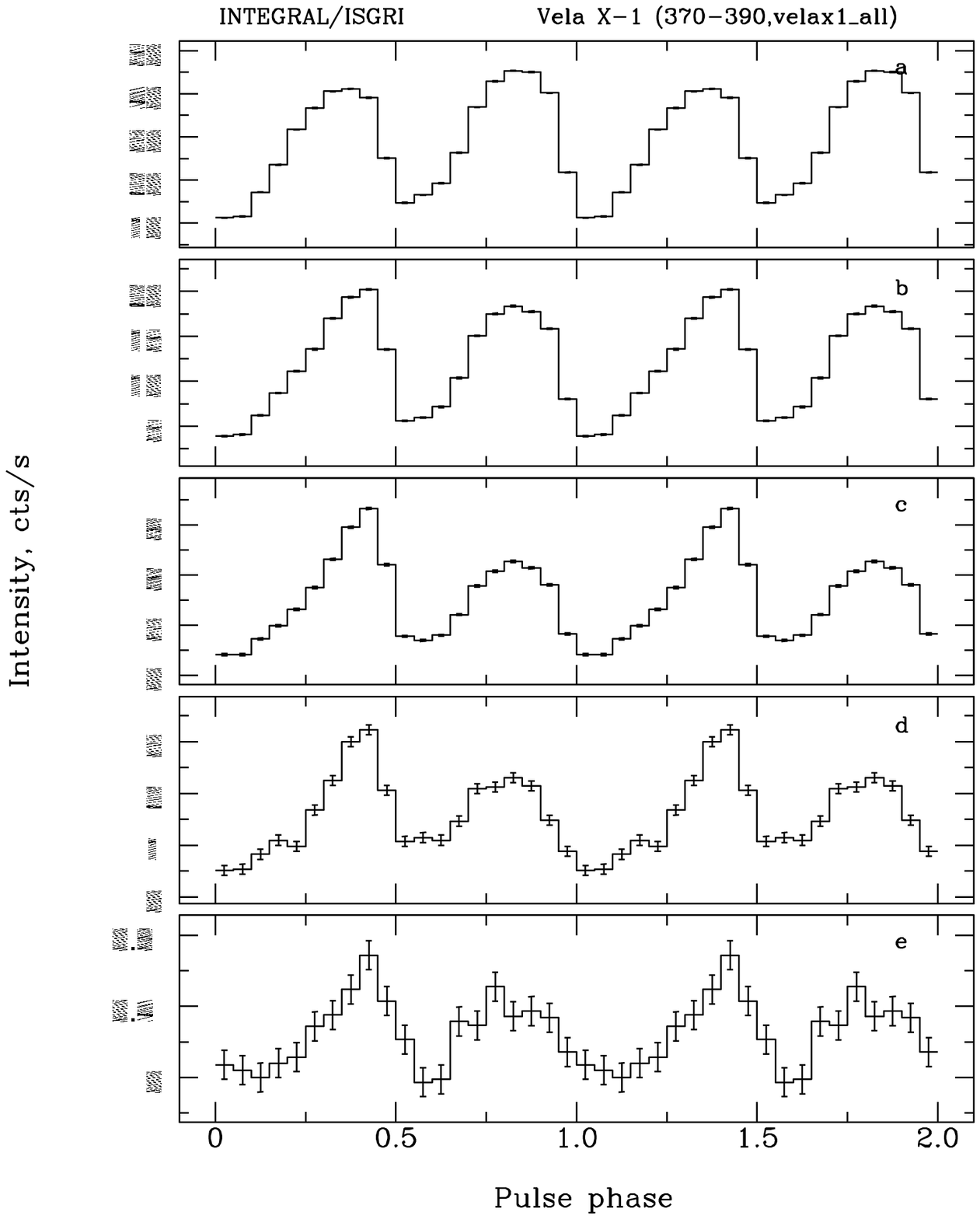}
}
\hbox{
  \includegraphics[width=0.5\textwidth]{./figures/herx1_high_339_339_rev_contour_col.ps2}
  \includegraphics[width=0.5\textwidth]{./figures/velax1_all_370_390_rev_contour_col.ps2}
}
}
  \caption{{\it Upper panels.} Her X-1 and Vela X-1 pulse profiles, obtained with the 
INTEGRAL observatory in several energy channels: 20-30 keV (a), 30-40 keV (b), 
40-50 keV (c), 50-70 keV (d) and 70-100 kev (e). {\it Bottom panels.} Correspontind 
relative intensity maps. Dashed lines denote positions of harmonics of the cyclotron 
line.}\label{pp_maps}
\end{figure}

In the analysis of pulse profiles we also widely applied the intensity maps,
constructed in a 2-dimensional plane: energy and pulse phase. This method
suggested by Tsygankov et al.\cite{tsy06} allows one to trace large-scale
variations and ratios of peaks in the pulse profile with the energy and
pulse phase. In Fig.\ref{pp_maps} the pulse profiles in several energy bands
and corresponding intensity maps are shown for X-ray pulsars Her X-1 and
Vela X-1. For Vela X-1, it is interesting to note the increase of the
relative intensity of a second peak near 56 keV, where the harmonic of the
cyclotron line is registered; also, it is clearly seen that peaks of the
Vela X-1 pulse profile come together with the energy.

At the moment a number of pulse profiles and corresponding relative
intensity maps are constructed and studied for all states of bright X-ray
pulsars which were observed with the INTEGRAL observatory (\cite{tsylut08}).

\subsection{Pulse Fraction}

\begin{figure}
  \includegraphics[width=0.9\textwidth,bb=40 245 525 805,clip]{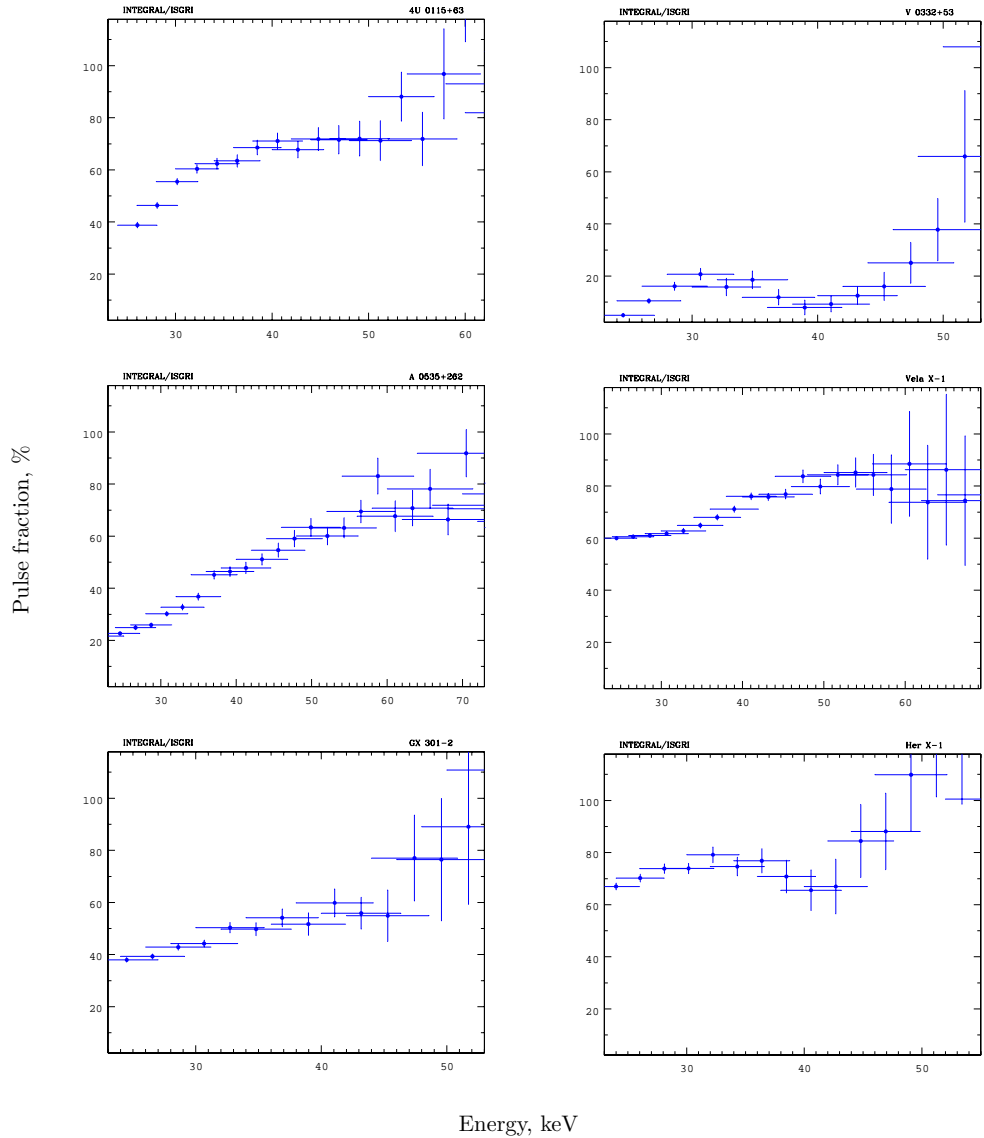}
  \caption{Pulse fraction dependence on the energy}\label{ppf_ener}

\end{figure}
\begin{figure}
  \includegraphics[width=0.9\textwidth,bb=40 245 525 805,clip]{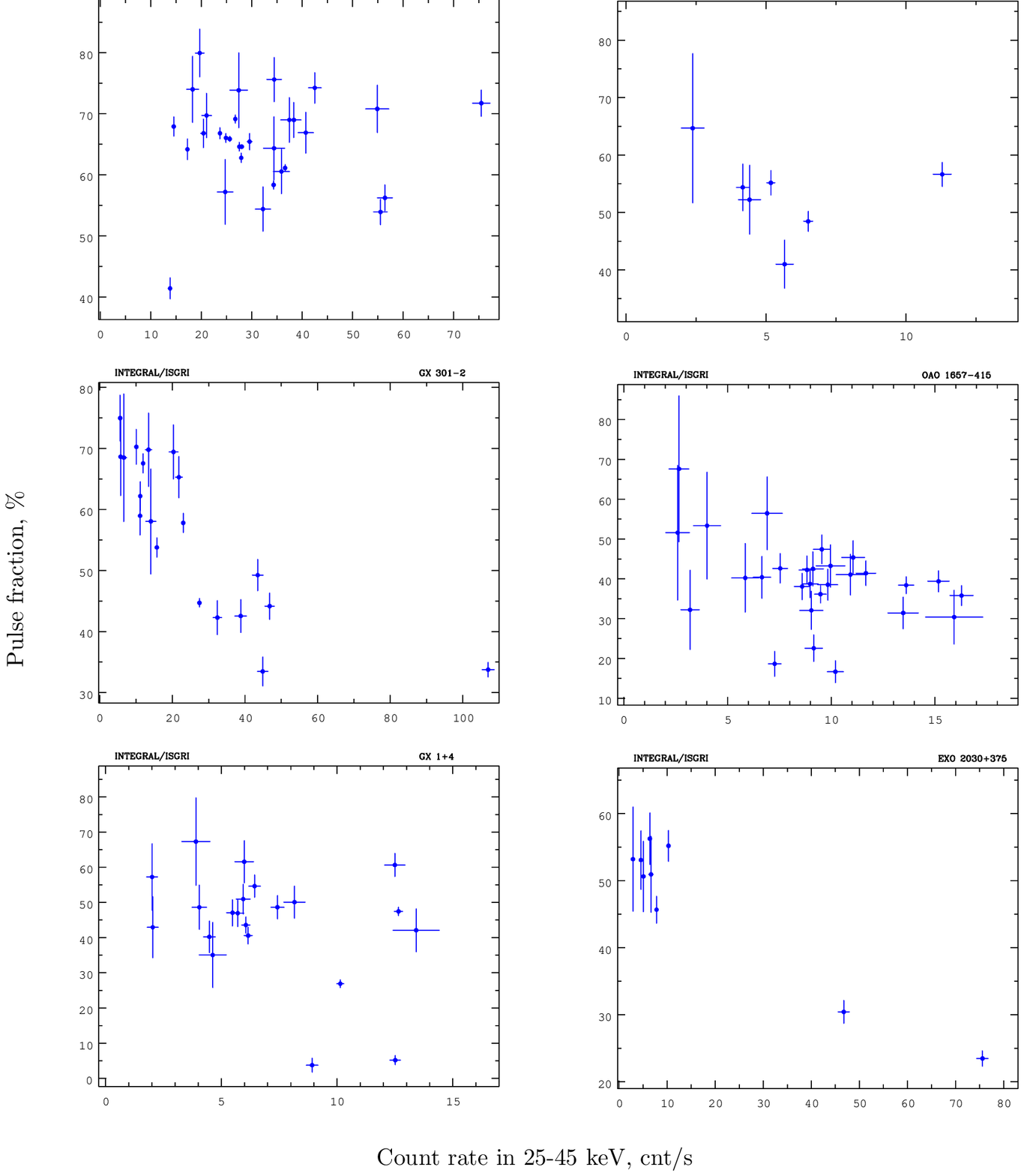}
  \caption{Pulse fraction dependence on the source intensity}\label{ppf_lum}
\end{figure}

The important and still poorly studied observational characteristic of the
X-ray pulsars radiation is the pulse fraction and its dependence on the
energy and luminosity.  Firstly, it was shown for 4U0115+63 that the pulse
fraction increases both with decreasing of the intrinsic source luminosity
and increasing energy \cite{tsy07}. This result can be qualitatively
explained and understood in terms of the ''toy'' model suggested above. As
the luminosity rises, the geometrical sizes of the emitting regions increase
and, accordingly, the pulsations are "smeared". The increase of the pulse
fraction with energy can also be explained by the fact that the emitting
regions are become more compact. In the same work \cite{tsy07} it was shown
that the growth of the pulse fraction with energy is not homogeneous -- near
the cyclotron energy and its harmonics a significant increase of the pulse
fraction was revealed both with INTEGRAL and RXTE observatories. The value
of the increase is near proportional to the depth of the cyclotron
line. This nontrivial effect was confirmed now for several other X-ray
pulsars (see \cite{tsylut08} and Fig.\ref{ppf_ener}).  The rise of the
pulsation amplitude with energy can be connected with the shape of X-ray
pulsars' spectra, as the intensity in the Wien tails will be strongly
modulated. The large modulation of the signal in the cyclotron line can be
caused by the strong dependence of the opacity on the angle with the
magnetic field.

As mentioned above the pulse profiles offen depend on the source intensity,
which can lead to variations of the pulse fraction. To check this assumption
we also built the pulse fraction dependence on the source intensity in the
25-45 keV energy band, where source fluxes are maximum
(Fig.\ref{ppf_lum}). For several sources (GX301-2, possibly OAO 1657-415,
EXO 2030+375) the pulse fraction is decreased with the growth of the source
intensity, which can be understood and explained in terms of the above
''toy'' model, when the increase of the accretion rate cause an increase of
the accretion column height. Another group of sources is the X-ray pulsars,
for which a significant scatter of pulse fraction values is observed for
close luminosities, but the general trend in a wide dynamical range of
observed fluxes is absent (Vela X-1, Cen X-3, GX1+4). In this case, most
probably, the source luminosity is not enough for the formation of accretion
columns and the emission is formed near the neutron star surface (this is
cicumstantially confirmed by the absence of the cyclotron line energy
changes with the luminosity for these sources). The observed scatter of
pulse fraction values is possibly connected with local inhomogeneities of
the stellar wind or accretion flows.

\section{Summary}

\begin{itemize}

\item Most accurate results of the spectral and timing analysis of several
dozens X-ray pulsars in hard X-rays are obtained at the moment with the
INTEGRAL and RXTE observatories.

\item The evolution of the cyclotron energy with the source luminosity was
studied in detail for the first time. It was shown that for V0332+53 this
dependence is linear (the cyclotron line energy is increased with the
luminosity decrease), but for 4U0115+63 and A0535+262 it is more
complicated; the ''reverse'' behaviour is detected for Her X-1 and GX301-2.

\item The strong dependence of the pulse fraction on the energy and source
luminosity is revealed and study in detail.

\item The prominent feature in the dependence of the pulse fraction on
energy band was revealed near the cyclotron frequency for several bright
sources.

\item There are preliminary ideas how to explain the above, but the
additional theoretical studies and simulations are needed for resolving the
puzzles.

\end{itemize}


\begin{theacknowledgments} 

We thank J.Poutanen, M.Revnivtsev and V.Suleimanov for the useful discussion
during meetings in the ISSI (Bern), whose support and hospitality we also
acknowledge. This work was supported by the Russian Foundation for Basic
Research (Projects No. 07-02-01051 and 08-08-13734), the Presidium of the
Russian Academy of Sciences (The Origin and Evolution of Stars and Galaxies
Program), and the Program of the President of the Russian Federation
(Project No. NSh-5579.2008.2).

\end{theacknowledgments}



\begin{thebibliography}{99}

\bibitem{wangwel81}
Wang Y.-M., Welter G. \emph{Astron. Astrophys.} \textbf{102}, 97
(1981)

\bibitem{white83}
White N., Swank J., Holt S. \emph{Astrophys. J.} \textbf{270}, 771
(1983)

\bibitem{bs1976}
Basko M.M., Sunyaev R.A.  \emph{Astron. Astrophys.} \textbf{42}, 311
(1975)

\bibitem{mesnag81}
Meszaros P., Nagel W. \emph{Astrophys. J.} \textbf{299}, 138 (1981)

\bibitem{tsy06} 
Tsygankov S., Lutovinov A., Churazov E., Sunyaev R. \emph{MNRAS} 
\textbf{371}, 19 (2006)

\bibitem{tsy07} 
Tsygankov S., Lutovinov A., Churazov E., Sunyaev R. \emph{Astronomy Letters}	 
\textbf{33}, 368 (2007)

\bibitem{fer07}
Ferrigno C., Segreto A., Santangelo A., et al. \emph{Astron. Astrophys.} 
\textbf{462}, 995 (2007)

\bibitem{barn08} 
Barnstedt J., Staubert R., Santangelo A., et al. \emph{arXiv:0805.1811} (2008)

\bibitem{tsylut08} 
Tsygankov S., Lutovinov A. \emph{Astronomy Letters}, \textbf{34}, in press
(2008)

\bibitem{win03}
Winkler C., Courvoisier T., Di Cocco G., et al. \emph{Astron. Astrophys.} 
\textbf{411}, L1 (2003)

\bibitem{fil05} 
Filippova E., Tsygankov S., Lutovinov A., Sunyaev R. \emph{Astronomy
Letters} \textbf{31}, 729 (2007)

\bibitem{lut04} 
Lutovinov A., Tsygankov S., Revnivtsev M., et al.  \emph{The INTEGRAL
Universe}, edited by V. Sch\"onfelder, G. Lichti \& C. Winkler, Proceedings
of the 5th INTEGRAL Workshop, ESA-\textbf{552}, 253 (2004)

\bibitem{mih98} 
Mihara T., Makishima K., Nagase F. \emph{Adv. Space Res.} \textbf{22}, 987
(1998)

\bibitem{cab08}
Caballero I., Santangelo A., Kretschmar P., et al. \emph{Astron. Astrophys.} 
\textbf{480}, L17 (2008)

\bibitem{staub07}
Shtaubert R., Shakura N., Postnov K., et al. \emph{Astron. Astrophys.} 
\textbf{465}, L25 (2007)

\bibitem{ghosh79}
Ghosh P., Lamb F. \emph{Astrophys. J.} \textbf{234}, 296 (1979)

\bibitem{pacheco87}
Janot-Pacheco E., Motch C., Mouchet M. \emph{Astron. Astrophys.} 
\textbf{177}, 91 (1987)







\end{thebibliography}
\end{document}